\documentclass[fleqn,10pt]{wlscirep}
\usepackage[utf8]{inputenc}
\usepackage[T1]{fontenc}
\title{2D tunable all-solid-state random laser in the visible.}

\author[1]{Bhupesh Kumar}
\author[1]{Ran Homri}
\author[1,*]{Patrick Sebbah}
\affil[1]{Department of Physics, The Jack and Pearl Resnick Institute for Advanced Technology, Bar-Ilan University, Ramat-Gan, 5290002 Israel.}

\affil[*]{patrick.sebbah@biu.ac.il}

\begin{abstract}
A two-dimensional (2D) solid-state random laser emitting in the visible is demonstrated, in which optical feedback is provided by a controlled disordered arrangement of air-holes in a dye-doped polymer film. We find an optimal scatterer density for which threshold is minimum and scattering is the strongest. We show that the laser emission can be red-shifted by either decreasing scatterer density or increasing pump area. We show that spatial coherence is easily controlled by varying pump area. Such a 2D random laser provides with a compact on-chip tunable laser source and a unique platform to explore non-Hermitian photonics in the visible.
\end{abstract}
\begin{document}

\flushbottom
\maketitle
%
%
\thispagestyle{empty}


\section*{Introduction}
Conventional lasers are known for their high spatial coherence due to their limited number of spatial modes. Because of this property, laser emission is highly directional. On the other hand, high spatial coherence causes speckle noise \cite{goodman}, which is an undesired effect in laser-based display systems, as it destroys the information content and reduces resolution \cite{katz96}. 
Random lasers (RL) are a new class of lasers, where light is confined by multiple elastic scattering in an active disordered medium \cite{Cao99}. RL are an interesting alternative to conventional lasers as they can provide low spatial coherence due to radiation of several uncorrelated lasing modes \cite{redding12}.
Such laser devices can prevent the formation of speckle and produce high-quality images similar to those produced by conventional spatially incoherent sources such as light-emitting diodes \cite{polymer}.  
Other potential applications of RL can be found in display lighting \cite{kishore2010}, document encoding \cite{rmbala96}, bioimaging \cite{redding12} tumor detection \cite{polson04} or sensing \cite{ismail15}. Solid-state dye lasers have long been foreseen as an alternative in the visible to the toxic and complex dye lasers, leading to highly-efficient tunable lasers, with potential applications in integrated photonics \cite{Costela09}. Low-cost and flexible random lasers have been proposed by introducing random scattering in dye-doped organic films in various ways \cite{Sznitko15}, including by rubbing surface \cite{Sznitko13}, using biological living Cells \cite{zhiyang22}, by laser ablation \cite{Consoli19}, or by introducing nanorods \cite{Wang16, GeK21}, polycrystalline thin film\cite{Zong20},  nematic liquid crystals \cite{Dai20}, nanosheets \cite{Cerdan12}, dye aggregates \cite{Ye17}, dielectric \cite{Zhao11}, ZnO \cite{Haider19}, or metallic nanoparticles \cite{Popov06}.
In all these cases however, the scattering medium is not controllable, resulting in aggregation of scattering particles and nanowires that leads to unpredictable effective scatterer size and distribution.
 To date, two-dimensional (2D) planar waveguide random lasing with deterministic disorder has been demonstrated for non-visible wavelengths including Terahertz range \cite{zeng18,zeng16,rotter16}, and mid-infrared regime \cite{yzhang13,seassal19}. 
Recently, we reported deterministic disorder-based one-dimensional all-solid-state random laser, where sub-micrometric grooves were carved in a  thin polymer layer using e-beam lithography \cite{Kumar21}. In this letter, we extend this method to design a 2D random laser in dye-doped polymer films with a deterministic disordered distribution of air-holes. Such a 2D well-controlled structure offers the possibility to investigate new aspects of random lasing without any limitation due to sample optical damage. Here, we observe sharp lasing peaks under uniform optical pumping. We confirm that lasing results from multiple scattering by showing how the emission spectrum is sensitive to local pumping. Lasing threshold is measured as a function of scatterer density and pump area. An optimal density is found where lasing threshold is minimal and scattering is the strongest, while control of the spatial coherence is achieved by simply varying the beam diameter. Finally, we show that emission spectrum can be tuned by varying the scatterer density as well as the pump area.
Tailoring the disorder in random lasers dye-doped polymer films opens new perspectives to investigate non-Hermitian optics, including modal signatures of 2D random lasers, role of spatial correlations \cite{Lin20}, impact of local perturbation and exceptional points \cite{Bachelard22}, and control of laser characteristics \cite{Rotter13}.
 Incoherent light with a broad spectrum of several tens of nanometers can be used for applications based on low coherence interferometry such as optical coherence tomography\cite{karamata04}.
\\

\section*{Methods}
\subsection*{Sample Preparation}
\begin{figure}[ht!]   
\centering
\includegraphics[width=\linewidth]{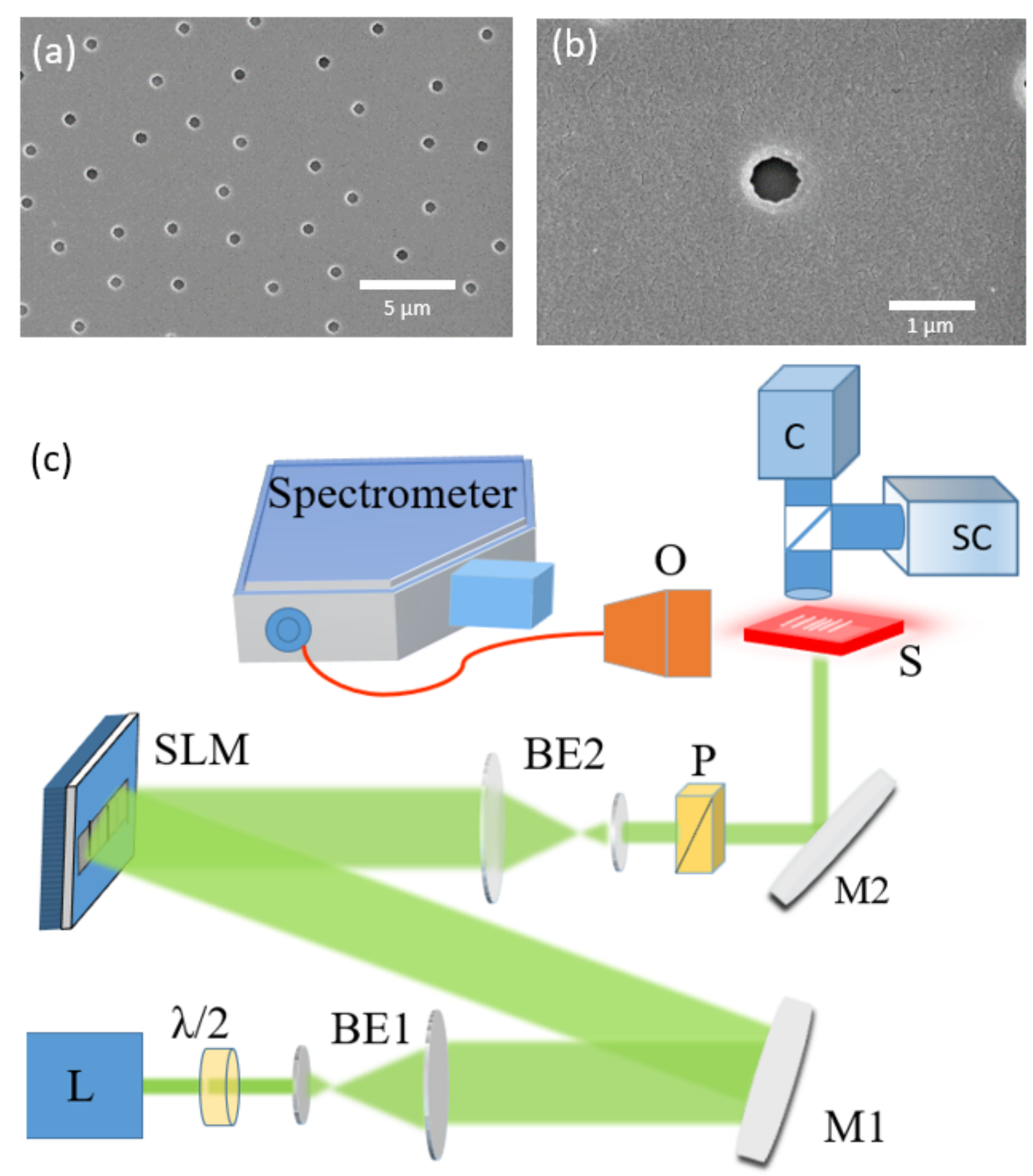}
\caption{\textbf{Photonic Structure and Experimental Setup:} 
(a) Scanning electron microscope image of a section of the disorder sample with surface filling fraction 1.1\%. (b) Zoom-in top view of the etched 500 nm-diameter air holes. (c) Schematic of the experimental set-up: SLM: Spatial Light Modulator; L: Laser; M1 and M2: Mirrors; BE1 and BE2: Beam expanders; P: Polarizer; SC: Streak camera; C: sCMOS Camera; O: Microscope Objective.}
\end{figure}

To fabricate our 2D solid-state random laser, we use PMMA (polymethyl methacrylate, from Microchem,USA) polymer with a molecular weight of 495000 g/mol at a concentration of 6\% weight in anisole. It is doped with 5\% weight of DCM (from Exciton) laser dye (4-dicyanomethylene-2-methyl-6-(4-dimethylaminostyryl-4H-pyran). DCM dye is preferred because it has a fluorescence spectrum centered around 600 nm with a good quantum yield and a large Stokes shift (100 nm), which reduces reabsorption of the emitted light. 

A 600 nm PMMA-DCM polymer layer is deposited on a fused silica substrate (Edmund Optics) using spin coating (1000 rpm, 60 s) \cite{Kumar21}. The obtained PMMA-DCM layer is annealed at 120 ºC in an oven for 2 hours to induce the polymerization process. The index of refraction of this layer is $n$=1.54. A Matlab code has been developed to generate the 2D disordered patterns for a given surface filling fraction (FF) of holes, positioning them on a circular surface of diameter D = 1200 $\mu$m, using a uniform probability distribution (Mersenne Twister pseudo-random number generator). The algorithm forces a minimum edge-to-edge distance of 2 $\mu$m between each pair of holes to avoid proximity effects during the etching process.  E-beam lithography in raster-scan mode is used to carve the photonic structure on the planar waveguide. To avoid charge accumulation during the lithography process, samples are coated with a conductive polymer (Espacer) of thickness 40 nm, removed after exposure by immersing the sample in deionized water for 40 sec. A set of 2D disorder samples has been prepared, having surface filling fraction FF = 0.17 \%, 0.35 \%, 0.70 \%, 1.1 \%, 1.4 \%, 1.9 \%, and 2.4 \%. All samples are prepared under identical conditions. 
The diameter of the holes is 500 nm, to enhance scattering in the vicinity of the first Mie resonance at emission wavelengths. 

From high resolution SEM images of a section of the structure, as shown in Fig.1(a,b). The fabrication method ensures a refractive index contrast of 0.54 between air holes and the polymer layer. 

\subsection*{Experimental Setup}
A frequency doubled mode-locked Nd:YAG laser ($\lambda$ = 532 nm, maximum output energy 28 mJ, pulse duration: 20 ps, repetition rate: 10 Hz, PL2230 Ekspla) is used to optically pump the sample at a repetition rate of 10 Hz. 
The pump beam is first expanded 5$\times$ to be spatially modulated by a 1952×1088-pixels reflective spatial light modulator (SLM) (HES 6001 from Holoeye, pixel size 8.0 $\mu$m). The SLM is used here in intensity modulation mode. The pump beam is circularly-shaped by the SLM before being imaged on the sample surface using a 4 $\times$ reducing telescope. Top surface of the sample is imaged using a sCMOS camera (Zyla 4.2 from Andor, 22 mm diagonal view, 6.5 $\mu$m pixel size) placed at the top of a fixed-stage microscope (AxioExaminer A1 from Zeiss). A 532 nm notch filter (NF533-17 from Thorlabs) is used to remove residual pump light. Computer-generated patterns sent to the SLM are used to control the pump size and pump energy (0-255 grey scale level) delivered to the sample. Pump energy reaching the sample is monitored using a photodiode laser energy sensor (PD10-C from Ophir). 
Lasing emission is collected in the x-y plane by a microscope objective (20x from Thorlabs) connected to a high-resolution imaging spectrometer (iHR550 from Horiba) via a multimode fiber. 
The spectrometer is equipped with a 2400 $mm^{-1}$ density grating,  which provides a spectral resolution of 20 pm. 
Even though DCM dye molecules are distributed isotropically inside the PMMA polymer matrix, in-plane lasing emission is not uniform due to fluorescence anisotropy \cite{lebental12}. We therefore perform all measurements in the direction of maximum emission.
For time-domain measurements, out-of-plane light scattered from air holes after passing through the notch filter @532 nm is directed toward a custom-designed streak camera (AXIS-QVM). The temporal resolution of the camera is 3 ps for single shot measurement. The experimental setup is illustrated schematically in Fig.1(c).\\
\begin{figure}[ht!]   
\centering

\includegraphics[width=\linewidth]{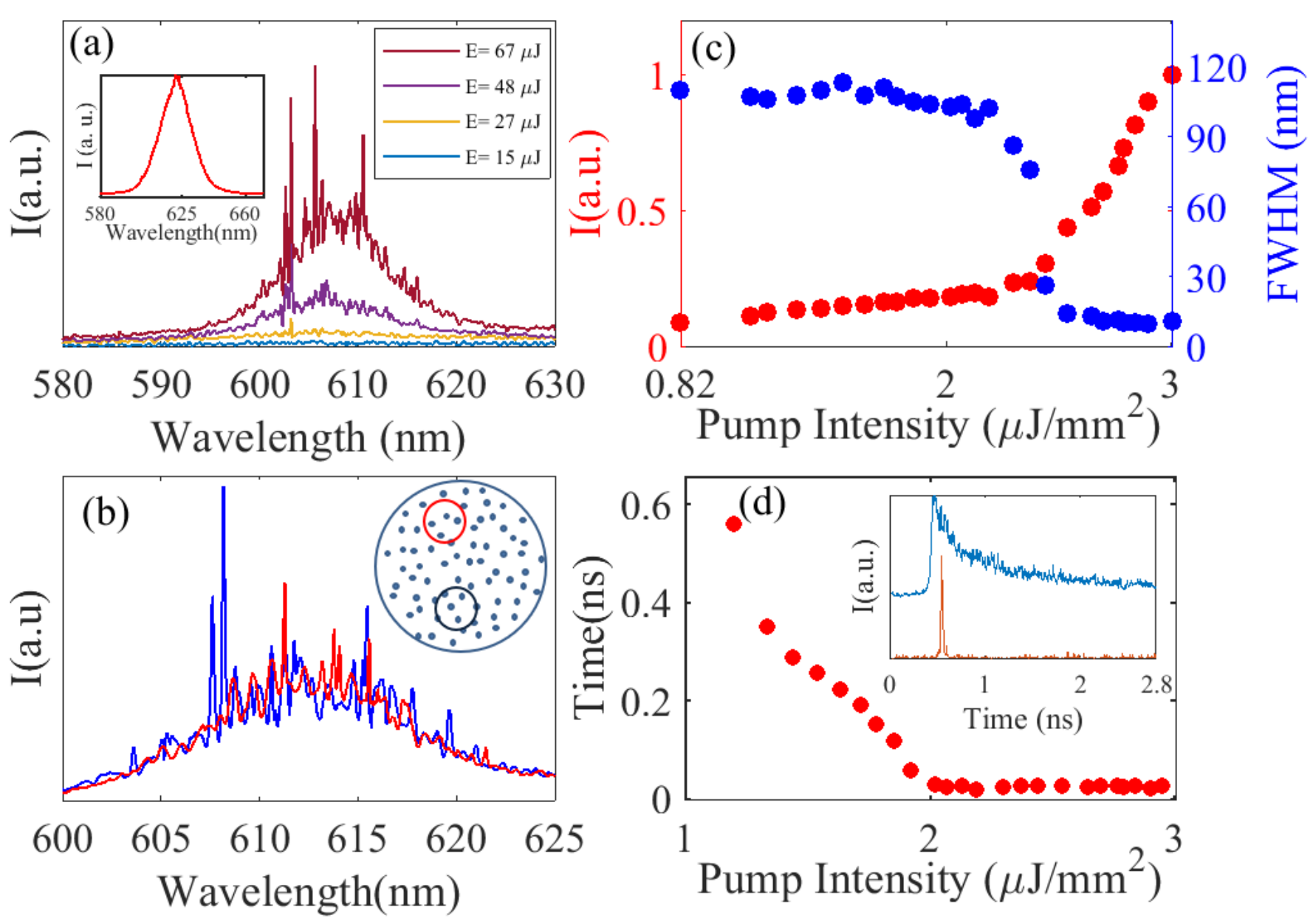}
\caption{\textbf{Laser Characteristics:} 
(a) Emission spectrum measured for 4 different values of pump energy density (FF = 1.40 \%, pump diameter 100 $\mu$m). Inset shows photoluminescence spectrum of DCM dye in PMMA polymer matrix, in the absence of air holes. (b)  Emission spectrum measured at 2 different positions (FF = 1.1 \% and pump diameter 100 $\mu$m). inset: pump locations on the sample surface (density of scatterers not to scale). (c) Integrated intensity and full width at half maximum (FWHM) of emission spectrum plotted as a function of pump energy density. (d) FWHM of the measured temporal profile with increasing pump energy density. Inset: temporal profile measured below (blue) and above threshold (red).}
\end{figure}
\section*{Results}
We first investigate the laser characteristics of the sample with filling fraction, FF = 1.40 \%, when optically pumped at $\lambda$ = 532 nm, with a 100 $\mu$m-diameter circular beam. Spectra have been averaged over 10 shots. Peak position does not change from shot to shot. At low excitation energy, the sample emits broad spontaneous emission.  When the pump energy exceeds some lasing threshold, discrete ultra-narrow linewidth peaks emerge in the emission spectrum, as shown in Fig.2(a). Full-width at half-maximum (FWHM) of individual lasing mode is 0.2 nm, which is the resolution limit of the spectrometer, and is orders of magnitude smaller than the 21 nm linewidth of spontaneous emission measured below threshold. Observation of ultra-narrow spectral features within the gain peak of the DCM dye is a clear indication of coherent laser oscillations.
Emission spectra measured for two different pump locations on the sample are shown in Fig.2(b). Because different regions of the disordered structure are probed, coherent excitation of different modes associated with different disorder configurations is observed. As a result, random positioning of the pump will generate random emission spectra. Although the position of the holes is determined by design and the structure itself is not strictly speaking random, nevertheless, the spectral dependence with pumping region is the signature of the random nature of this lasing device. This dependence of the laser spectrum with pump position has recently been availed to control  random laser emission by shaping spatially the pump intensity profile \cite{leonetti11, patrick14,Kumar21}.
\begin{figure}[ht!]   
 \centering

\includegraphics[width=\linewidth]{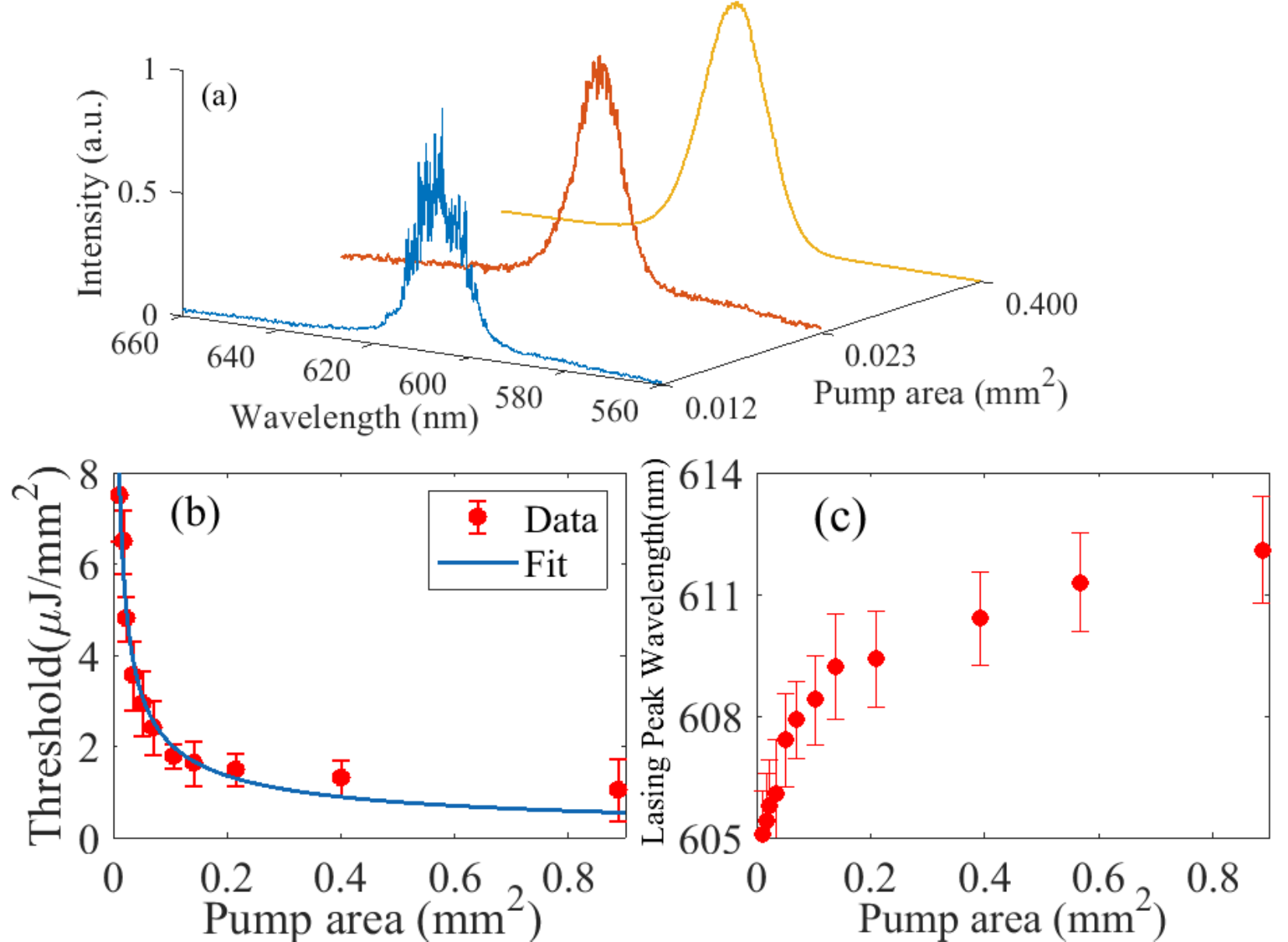}
\caption{\textbf{Threshold, emission wavelength, lasing modes vs. Pump Area:} (a) Normalized emission spectrum recorded for a pump areas of 0.012 $mm^{2}$, 0.023 $mm^{2}$ and 0.400 $mm^{2}$  (b) Pump energy density at lasing threshold, $I_{th}$, is plotted as a function of pump area, $A$, for a sample having filling fraction FF= 0.17 $\%$. The data has been fitted by a power law, $I_{th}$ = $0.005/A^{0.57}$. (c) Central peak of the emission spectrum plotted vs. pump area.}
\end{figure}
The integrated intensity and full width at half-maximum (FWHM) of the emission spectrum recorded for a pump diameter of 1000 $\mu$m are shown in Fig.~2(c) as a function of the incident pump energy density for a sample with FF = 1.1 \%. Lasing threshold behaviour is confirmed by the change in slope of the integrated intensity and the rapid fall of the FWHM of the emission spectrum. This occurs for this particular sample at a pump energy density of 2.3 $\pm$0.23  $\mu$J/$mm^{2}$. 

The temporal pulse profile of the laser emission is measured with the streak camera for increasing pump power energy. Inset in Fig.2(d) shows the temporal profile measured for pump intensity below and above the threshold. When pump power is below the threshold, dye molecules experience spontaneous decay, a relatively slow process, which results in a slow exponential fall-off (FWHM = 0.20 ns). When the system is pumped above the threshold, the pulse profile narrows down dramatically and becomes similar to the profile of the pump pulse (FWHM = 20 ps). Threshold behavior is characterized in the time domain by plotting the pulse width as a function of increasing pump energy density (Fig.2(d)). Inflection point occurs at a pump energy density of 2.1 $\pm$0.32  $\mu$J/$mm^{2}$, which is within the error bars of the threshold value found in Fig. 2(c).\\
Earlier studies show that random laser spatial coherence can be controlled by adjusting the scattering strength and the pump geometry \cite{cao11}. Fundamentally, increasing the number of uncorrelated lasing modes reduces the spatial coherence of the RL. This is simply achieved in our case by increasing the pump area. Figure 3(a) shows emission spectra recorded for three different pump areas of 0.012 $mm^{2}$, 0.023 $mm^{2}$ and 0.400 $mm^{2}$.

For small pump area, lasing peaks can be resolved spectrally as seen in Fig.3(a)(blue plot). For larger pump diameter (>250µm), modal density becomes large and lasing peaks strongly overlap spectrally (brown plot fig.3(c)). On further increasing pump area, the emission spectrum appears quasi-continuous because of high lasing mode density. This shows the ability to control the spatial coherence of this device by simply varying the beam diameter.
Next, we investigate the dependence of the lasing threshold on pump area for a sample having FF = 0.17 \%. Laser threshold versus pump spot diameter is shown in Fig. 3(b). The pump spot diameter, which varies between 120 $\mu$m and 1000 $\mu$m, is directly measured by imaging the sample surface from the top. The laser threshold is measured by plotting the maximum of the emission spectrum as a function of pump intensity. For small pump diameter where emission spectra have discrete multiple lasing peaks (blue plot in Fig. 3(b)), we consider the intensity count of the highest peak, which is near the center of the emission spectra. For larger pump diameter, when mode overlap is strong (brown plot in Fig. 3(b)), the maximum of the global emission spectrum is considered.
We find experimental data are well fitted by a power law, with an exponent $-0.57$. This is consistent with values reported in the literature \cite{hcao_pra}. Power law dependence of the lasing threshold on pump area is attributed to the decrease in return probability of the scattered light in gain volume with decreasing pump size \cite{hcao_pra, gracia11}. Interestingly, we also observe a spectral shift in the emission spectrum peak when varying pump area. For a given pump energy density, a larger pump area emits a red-shifted emission spectrum compared to the emission spectrum at a smaller pump area. Fig.3(c) shows the redshift of the emission spectrum with increasing pump area. A total shift of 7 nm is observed when the pump diameter is varied from 100 $\mu$m to 1000 $\mu$m. 
We suggest a possible explanation for this observation. Out-of-plane scattering is the dominant loss mechanism in our system, since reabsorption by the DCM dye is negligible. Actually, this mechanism contributes to loss in two different ways, which must be balanced by gain \cite{yamilovPRL}: (a) “Vertical loss” in the gain region, which is uniform on average and does not affect the nature of the modes; (b) “Leakage” beyond the gain region, where photons are lost and never return in the gain region. This local loss induces modal confinement, modal intensity redistribution, and frequency shift \cite{yamilov05}. For large pumping area, the first loss mechanism dominates and is advantageously reduced by shifting to longer wavelengths, since out-of-plane vertical scattering decreases towards larger wavelengths. When pumping smaller area, leakage is naturally decreased by increasing scattering within the gain region. Increasing scattering is achieved by a shift towards smaller wavelengths (blueshift), where scattering is stronger.

We now investigate the role of air-hole density on random lasing. Lasing characteristics of samples with different filling fractions (FF = 0.17 \% to 2.43 \%) are measured for a pump diameter of 1000 $\mu$m. The spot diameter has been expanded to increase the number of modes to a point where spectral overlap smooths out the laser emission spectrum. Fig.4(a) shows that the lasing threshold first decreases with increasing filling fraction, then increases for sample with filling fraction above FF = 0.70 $\%$. \\
\begin{figure}[ht!]   
\centering

\includegraphics[width=\linewidth]{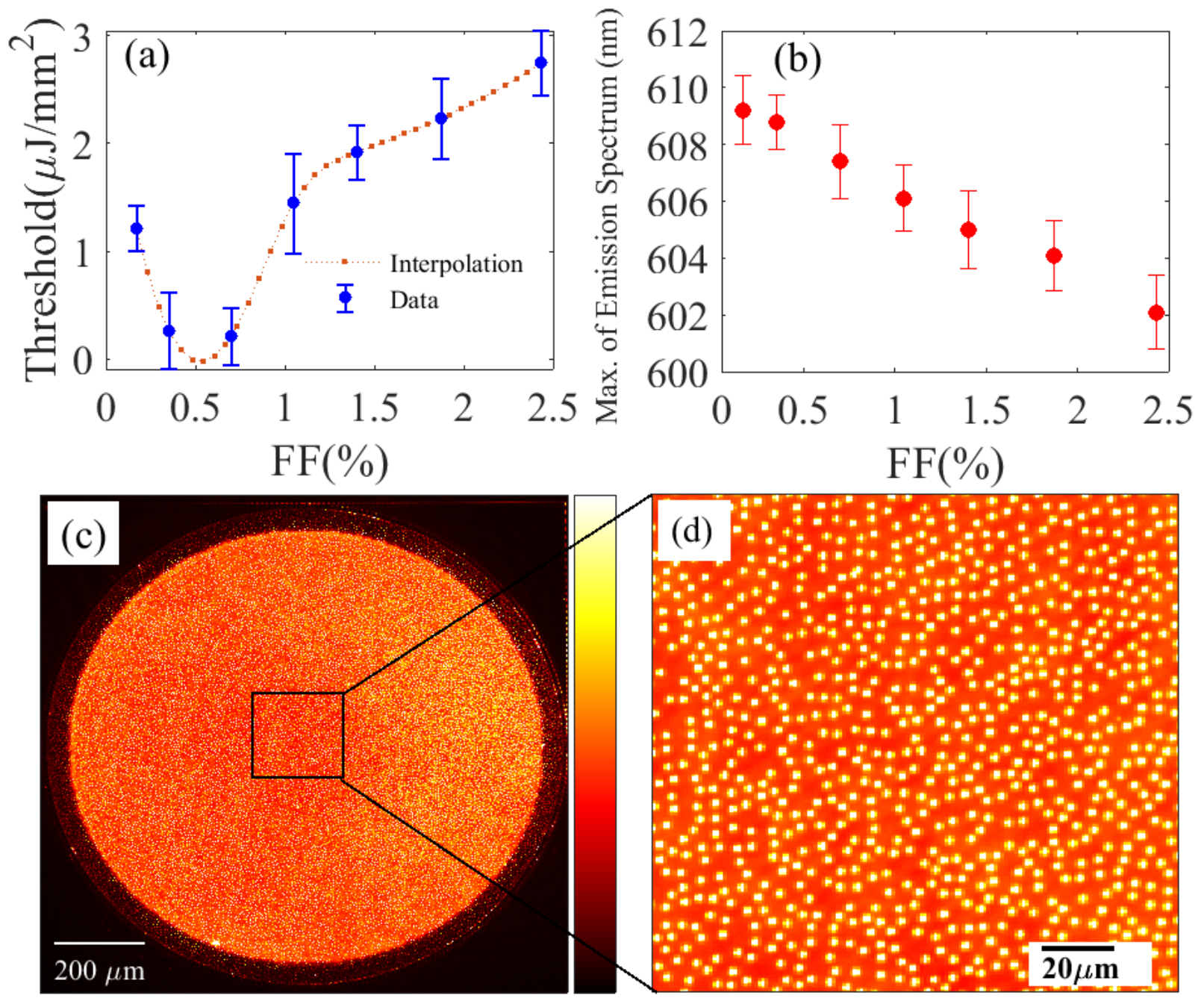}
\caption{\textbf{ Lasing Threshold and emission shift vs Filling Fraction:}
 (a) Lasing threshold is plotted for samples with different filling fractions (FF). (b) Measured central peaks of the 2D random laser emission spectra plotted for samples with different filling fraction ranging from 0.17 \% to 2.43 \%. All measurements are performed at a pump spot diameter of 1 mm. (c) Optical microscope image of lasing sample with FF= 1.4 \%. (d) Zoom-in on (c)}
\end{figure}
By interpolating the data, we can roughly estimate the optimal filling fraction, FF = 0.55 \%, corresponding to the minimum lasing threshold, which is reached when scattering is the strongest, modes are most confined and modal losses are minimal. 
We believe that the rapid drop observed in Fig.3(a) is a signature of the proximity to Anderson localization \cite{anderson58}. This is supported by estimating the mean free path, $\ell$, within the independent scattering approximation (IPA) \cite{ VandeHulst}, and the localization length, $\xi$, from a 2D diffusion-based theory \cite{gupta03}. For infinite air-cylinders (index 1) in dielectric material (index 1.54), at wavelength 608nm and for FF=0.55\%, we find $\ell$=466 nm and $\xi$=900 $\mu$m, which is comparable to the sample dimensions (1200 µm). Interestingly, random laser probes here the impact of dependent scattering on modal extension. Indeed, within the IPA which holds for low densities, the mean free path, which measures the scattering strength of the disordered medium scales as the inverse of the number density of scatterer \cite{ VandeHulst}. At higher densities however, IPA breaks down, interaction between scatterers can no longer be neglected and mean free path is expected to increase again \cite{ ping90}. Our measure of the threshold dependence with scatterer density allows us to identify the optimal density where multiple scattering is the strongest. We have also measured the photostability of our RL. We found that the integrated intensity measured in a sample with 0.70\% filling fraction pumped at 1 $\mu$J/$mm^2$ is reduced by half after 25000 pump laser pulses, which correspond to 41 minutes.

Another interesting outcome of this study is the observation of a blue shift of the emission spectrum when air-hole density is increased. 
Figure 4(b) shows a linear dependence of the spectral shift with increasing filling fraction. Interestingly, the random laser can be tuned over 10 nm. Here, we are only limited by high threshold and optical damage for filling fractions above FF=2.43 \% (see Fig. 4(a)). But in principle, the tunability range of the random laser can be readily increased by considering larger sample and larger pumped area.
We attribute this blue shift to the the decrease of the average refractive index with the increase of air-hole density. Indeed, a randomly-distributed cavity will be always resonant for the same wavelength, $\lambda_{res}=c/(n\times f_{res})$, independently of the refractive index, $n$, within the cavity. Therefore, the resonance frequency, $f_{res}$, must increase linearly with decreasing $n$ (increasing FF), which translate into the observed blue-shift. Figs.4(c,d)  show the optical microscope image at two different scales of the out-of-plane random laser emission from a sample with filling fraction 1.4 \%.

\section*{Discussion}
In conclusion, we have demonstrated and investigated random lasing action in the visible of a 2D polymer-based active planar structure with deterministic scattering disorder. Sharp distinct resonances are observed for small pump area, due to coherent multiple scattering. By changing the scatterer density, we find the optimal concentration which maximizes scattering and minimizes lasing threshold.  Influence of pump area on lasing characteristics has been investigated providing spectral tunability and controllable spatial coherence. This new RL device can be used to control
laser emission \cite{leonetti11,patrick14,schohuber20} as well as directionality \cite{Rotter13} using iterative pump shaping method. Near-field imaging of the top surface should allow to assess the regime of strong localization and modal confinement at the critical concentration \cite{Kumar21}. Such a device can be used to explore non-Hermitian photonic systems and can find applications in the production of novel 2D photonic functional materials with possible applications in bio-imaging.

\section*{Data availability}
The datasets generated and/or analysed during the current study are available from the corresponding author on reasonable request.

\section*{Acknowledgements}
We thank Profs. Hui Cao and Alexey Yamilov for useful discussions. We are grateful to Dr. Leonid Wolfson for his dedication to the lab, as well as Dr. Yossi Abulafia for his help in the fabrication process and the Bar-Ilan Institute of Nanotechnology and Advanced Materials for providing with fabrication facilities. This research was supported by  Israel Science Foundation (1871/15, 2074/15 and 2630/20); United States-Israel Binational Science Foundation (2015694 and 2021811). PBC Post-Doctoral Fellowship Program from the Israeli Council for Higher Education.

\section*{Author contributions statement}
B.K. fabricated all the samples.  B.K. conducted the experiment(s). R.H. wrote Matlab code to control pump size. B.K. and P.S. analyzed the results. PS and BK wrote the manuscript.  All authors reviewed the manuscript. 

\section*{Competing interests}
The authors declare no competing interests.
\end{document}